\documentclass{article}
\usepackage{spconf,amsmath,graphicx}
\usepackage{subfig}
\usepackage{multirow}
\usepackage[spaces,hyphens]{url}
\usepackage{array}
\usepackage{multirow}
\usepackage[table,x11names]{xcolor}
\usepackage{svg}
\usepackage{graphics}

\title{Image Segmentation Using Hybrid Representations}

\name{Alakh Desai$^1$, Ruchi Chauhan$^{1,2}$, Jayanthi Sivaswamy$^1$}
\address{$^1$Center for Visual Information Technology \\
        $^2$Center for Computational Natural Sciences \& Bioinformatics \\
        IIIT Hyderabad, India}

\begin{document}
%
\maketitle
\begin{abstract}
This work explores a hybrid approach to segmentation as an alternative to a purely data-driven approach. We introduce an end-to-end U-Net based network called \textit{DU-Net}, which uses additional frequency preserving features, namely the Scattering Coefficients (SC), for medical image segmentation. SC are translation invariant and Lipschitz continuous to deformations which help \textit{DU-Net} outperform other conventional CNN counterparts on four datasets and two segmentation tasks: Optic Disc and Optic Cup in color fundus images and fetal Head in ultrasound images. The proposed method shows remarkable improvement over the basic U-Net with performance competitive to state-of-the-art methods. The results indicate that it is possible to use a lighter network trained with fewer images (without any augmentation) to attain good segmentation results.
\end{abstract}
\section{Introduction}
\label{sec:intro} Medical image segmentation is a major yet difficult task in computer-aided diagnosis. It is a key requirement for obtaining diagnostic information, be it organs or lesions, assessing tumor size, treatment planning, and so forth. Medical images have specific challenges due to artifacts, different standards of image acquisition, limited resolution, limited availability of annotated data, poor signal to noise ratio, etc. \cite{segmentationChallenges}. Various methods have been devised to address segmentation ranging from active contours to graph search to registration to an atlas. More recently, segmentation is posed as a pixel-wise classification problem in the deep learning paradigm. U-net is the most popular deep CNN architecture proposed for segmentation. It is based on an encoder-decoder structure with skip connections at each resolution level. The U-net and its variations have been used for both 2D and 3D segmentation. \cite{UnethighlyCited} \cite{UnetRecent}


While a Deep Learning (DL) approach removes the need for handcrafted features, which can be both the bottleneck or source of error, being entirely data-driven also implies that the onus of feature learning is completely on the model. Despite the impressive success that the DL approach has achieved in solving many problems, some weaknesses remain: i) overfitting due to inadequate training data. Labeled data is scarce in the medical domain, more so for the task of image segmentation that requires painstaking delineations of boundaries; ii) the 'capacity' of a model depends on the depth and complexity of the neural network, which in turn causes an explosion of parameters. Deploying complex networks with millions of parameters on lightweight platforms is also an inhibitor for clinical adoption of solutions. A hybrid approach needs to be explored where some predefined features can be used to improve the performance and robustness of the model without the need for additional data, computational cost or expert help.

Mallat's exploration used a convolutional network with predefined wavelets as filters \cite{mallat2012group}, which helps derive representations at multiple scales. These were called Scattering Coefficients(SC) with useful properties such as being Lipschitz continuous to deformations and invariant to translations. Further, SC are stable and geometrically invariant, therefore more generic. SC has a strong mathematical basis, which may help to reduce the opacity of the purported ‘black-box’ models of deep learning. SC have been used as features in a more traditional setting (with SVM) \cite{Mallet1} used for digit recognition \& texture discrimination tasks and shown to achieve good results. On complex recognition tasks (on Pascal \& Caltech), however, SC could not surpass the performance of other supervised learning methods.

Other attempts to combine handcrafted and learned features have generally adopted fusion followed by SVM based classification approach. For instance, texture descriptors (local phase Quantization, Local Ternary Patterns, etc. ) were combined with learned (with a CNN) features in \cite{handcraftA} to train an SVM for image classification. The feature combination (rather than individual) was shown to improve the performance across different datasets. In this paper, we present a novel hybrid approach by combining SC and learned (encoded) features for the task of segmentation.

\section{Methodology}
\begin{figure}
 \centering
 \includegraphics[trim={0 1cm 0 2cm},width=0.5\textwidth,height=4.75cm]{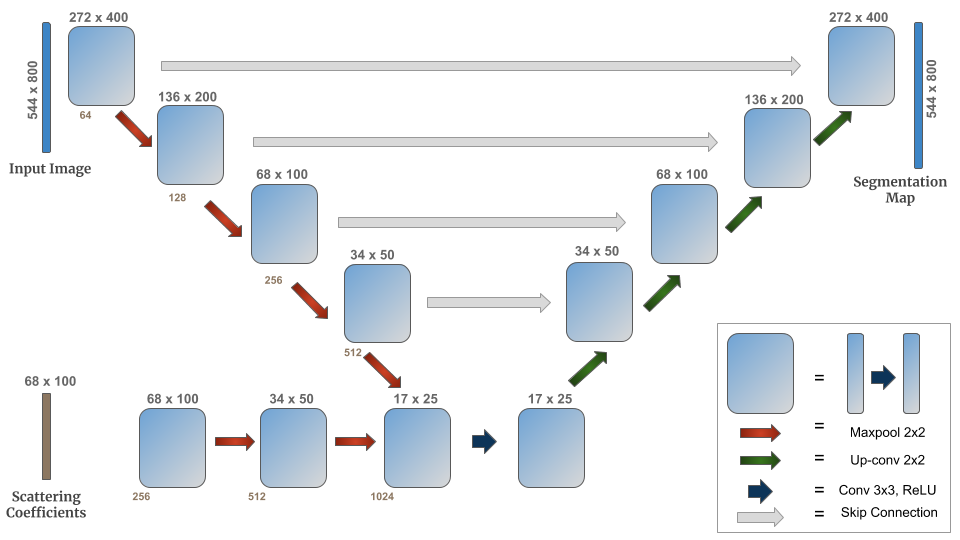}
  \caption{DU-Net Architecture}
  \label{DUNET}
\end{figure}

We propose an architecture called the \textit{DU-Net }, which is based on the U-Net. The inputs to this network are the given image and its SC. Separate encoders are designed for each of these inputs. The outputs of these encoders are concatenated and fed to a single decoder. 
In the image encoder, 3x3 convolutions were used along with element-wise ReLU for activation function; a 2x2 max pooling operation with stride two was used for down-sampling. The feature channels were doubled after each downsampling and halved after upsampling. The decoding units also use the features from image encoder through the skip connections. At each level, a skip connection was used to transfer the corresponding cropped feature map of the image and concatenate it to the up-sampled features to create the combined decoder feature map.  Finally, the high dimensional feature representation at the output of the final decoder layer is fed to the final convolutional layer with sigmoid activation to produce the pixel-wise probability map. The overall workflow is illustrated in Fig. \ref{DUNET}

The SC of the input image were derived using the ScatNet \cite{mallat2012group} with the Morlet wavelet. 
While the wavelet transforms are covariant to translations, the modulus pooling nonlinearity introduces translation invariance to the representations. The encoder for SC serves to do a feature reduction and consists of a series of convolution layers with elementwise ReLU. An average pooling was done after every second layer. 


\section{Experiments and Results}
\begin{figure} 
 \centering
  \includegraphics[scale=0.30]{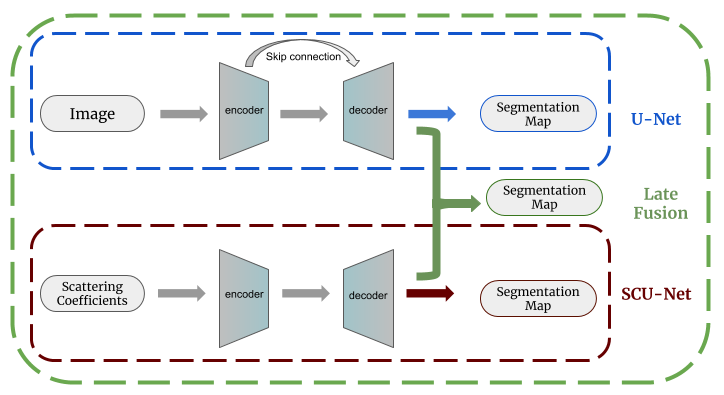}
  \captionsetup{justification=centering}
  \caption{Function Diagrams for the Variants \\ (UNet, SCU-Net \& Late Fusion)}
  \label{variants}
\end{figure}

In addition to the U-net, several variants of the \textit{DU-Net} were created for assessing the effectiveness of the proposed hybrid approach to segmentation. These are illustrated in Fig. \ref{variants} and described below. All implementations were done using the Keras framework on an Nvidia GTX 1080Ti GPU with 11 GB memory and  Intel Xeon E5-2640 processor.

\textit{Variant 1: SCU-Net.} This network is a version of the \textit{DU-net} with the image encoder removed.

 \textit{Variant 2: Late fusion Network.} In this architecture, a pair of encoder-decoder is used for each type of input: image and SC. The image encoder-decoder is similar to the U-Net, whereas SC-Decoder does not contain skip connections. The outputs are merged after the final level with a convolutional layer, which assigns the probability map. Late fusion combines the image and SC-based representations with a fixed weight whereas, \textit{DU-Net} takes their weighted combination.

 \begin{figure} 
 \centering
  \includegraphics[scale=0.4]{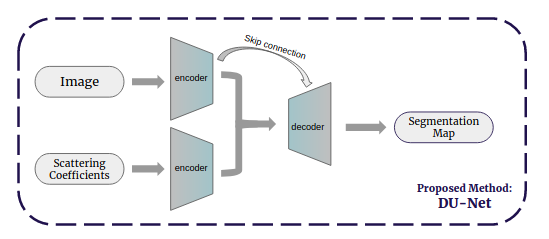}
  \caption{Function Diagram for DU-Net}
\end{figure}

{\renewcommand{\arraystretch}{1.495}
\begin{table*} 
\centering
\resizebox{13cm}{2cm}{
\begin{tabular}{|c|c|c|c|c|c|c|c|c|c|} 
\hline
\multirow{2}{*}{\textbf{Methods}} & \multicolumn{3}{c|}{ \textbf{Refuge}    }                                          & \multicolumn{3}{c|}{ \textbf{Dhristi-GS}    }                                      & \multicolumn{3}{c|}{ \textbf{RIM-One}    }                                          \\ 
\cline{2-10}
                                  & $\textbf{OC}_{\textbf{DICE (\%)}}$ & $\textbf{OD}_{\textbf{DICE (\%)}}$ & $\textbf{mIoU\textbf{(\%)}}$ & $\textbf{OC}_{\textbf{DICE (\%)}}$ & $\textbf{OD}_{\textbf{DICE (\%)}}$ & \textbf{mIoU(\%)} & $\textbf{OC}_{\textbf{DICE (\%)}}$ & $\textbf{OD}_{\textbf{DICE (\%)}}$ & \textbf{mIoU(\%)}  \\ 
\hline\hline
\textbf{FCN}\cite{FCN}                      & 84.67               & 92.56                        & 82.47                      & 87.95                        & 95.69                        & 83.92             & -                            & -                            & -                  \\ 
\hline
\textbf{U-Net}                    & 85.44               & 93.08                        & 83.12                      & 88.06                        & 96.43                        & 84.87             & -                            & 84.82                        & -                  \\ 
\hline
\textbf{M-Net}\cite{Mnet}                    & 86.48               & 93.59                        & 84.02                      & 88.60                        & 96.58                        & 85.88             & -                            & 86.51                        & -                  \\ 
\hline
\textbf{ET-Net}\cite{ETnet}                   & 89.12               & 95.29                        & 86.70                      & 93.14                        & \textbf{97.52}                        & 87.92             & -                            & -                            & -                  \\ 
\hline\hline
\textbf{SCU-Net}                  & 81.69               & 93.72                        & 84.61                      & 90.89                        & 95.22                        & 84.3              & 89.89                        & 90.96                        & 83.41              \\ 
\hline
\textbf{Late Fusion}              & 83.52               & 94.57                        & 85.08                      & 91.49                        & 96.89                        & 86.81             & 90.96                        & 92.3                         & 85.34              \\ 
\hline
\textbf{DU-Net}                   & \textbf{91.07}               & \textbf{96.42}                        & \textbf{88.78}                      & \textbf{96.01}                        & 97.11                        & \textbf{90.46}             & \textbf{93.24}                        & \textbf{96.16}                        & \textbf{87.39}              \\
\hline
\end{tabular}}
\caption{Performance on Optic Cup \& Optic Disk Segmentation}
\label{table:OCresults}
\end{table*}
}
To assess the utility of the proposed approach for different segmentation tasks experiments were done to extract the following: Optic cup, disk from color fundus images and the fetal head from ultrasound images. We next report the results of using different variants for each of these tasks.
\subsection{Optic Disc and Cup Segmentation}
In color fundus images, Optic disc (OD) is a bright yellowish elliptical area containing within, the optic cup (OC) region. Blood vessels and nerve fibers exit the retina from the OD region. OD segmentation involves challenges like peripapillary atrophy \& vessel occlusion whereas OC segmentation is even more demanding due to depth defined boundaries with indistinct gradient. The depth information is lost in the 2D fundus imaging process.

The proposed method has been evaluated on three publicly available datasets: DRISHTI-GS1 \cite{drishti}, RIM-ONE v2 \cite{RIM1}, and REFUGE. The OD region was localized using the Hough Transform. Vessels were detected using \cite{Vessel} and removed using diffusion-based inpainting \cite{Inpaint1}. The radius R was obtained from weighted circular Hough transform modeling OD as a bright, roughly circular structure that forms the base of the vessel tree. A region of interest of size 3R was extracted and resized to 512x512.

\begin{figure}
    \centering
    \includegraphics[scale=0.45]{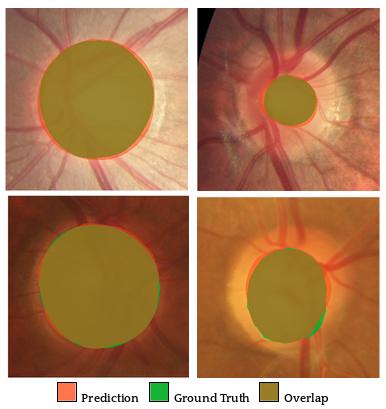}
    \caption{Qualitative results on fundus images. [Left: Optic Disk; Right: Optic Cup] }
    \label{fig:OCOD}
\end{figure}

For faster convergence and ensuring that OC is constrained inside OD, the network weights after training OD were used for OC instead of combined segmentation. The quantitative results using Dice Coefficient and mean Intersection over Union (mIoU) are presented in Table~\ref{table:OCresults}. The proposed \textit{DU-Net} is seen to outperform all existing methods across all datasets except for OD segmentation in the Drishti GS dataset. The state of the art method (\cite{ETnet}) uses augmentation with random changes in scale, rotation, \& colour jitter with an encoder pretrained on ImageNet. In contrast, the \textit{DU-Net} and its variants do not use any pretraining or augmentation. The qualitative results for OD and OC segmentation on a few sample test images are shown in Fig. \ref{fig:OCOD}

\subsection{Fetal Head Circumference}
Head Circumference (HC) is an important biometric to monitor the growth of the fetus. It is usually measured manually and thus subjected to inter and intra-observer variability. 

\begin{figure}
    \centering
    \includegraphics[scale=0.35]{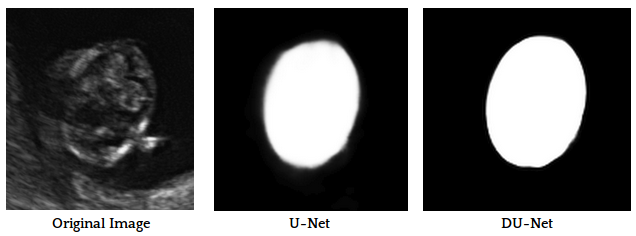}
    \caption{Probability maps from U-NET and \textit{DU-NET}}
    \label{fig:pm}
\end{figure}

The HC18 challenge dataset \cite{HC18paper} with 1334 images (of size ~ 540x800 px) was used. The ground truth (GT) is provided in the form of thin ellipses, which results in extreme class imbalance. Hence, a head mask was created with the given ellipse as the boundary. The computed head segment was post-processed to fit an ellipse for assessment \cite{ellipseFit}.  The given pixel to mm mapping was used to estimate the HC. The probability maps for U-Net \& \textit{DU-Net} are shown in the Fig \ref{fig:pm}. It can be observed that the map obtained from the U-Net has blurred edges while that of \textit{DU-Net} is sharp. The sharpness indicates the high confidence of the network while making predictions.

\begin{figure}
    \centering
    \includegraphics[scale=0.35]{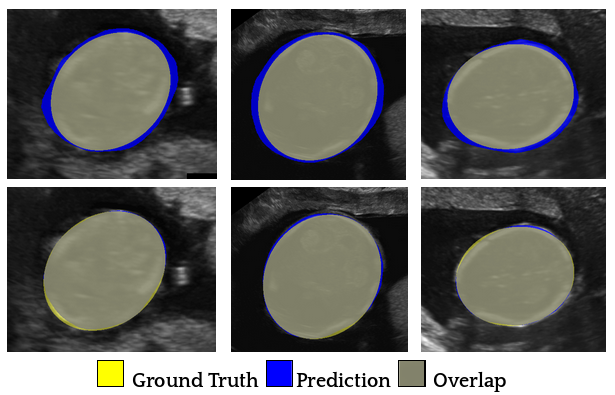}
    \caption{Sample segmentation results on Ultrasound images [Top: U-Net, Bottom: DU-Net]}
    \label{fig:US}
\end{figure}

\begin{table}
\centering \resizebox{6.8cm}{!}{  %
\begin{tabular}{|c|c|c|c|c|}
\hline 
\multirow{2}{*}{\textbf{Method} } & \multicolumn{2}{c|}{\textbf{Dice }} & \multicolumn{2}{c|}{\textbf{MHD (mm) }}\tabularnewline
\cline{2-5} 
 & \textbf{AVG}  & \textbf{STD}  & \textbf{AVG}  & \textbf{STD} \tabularnewline
\hline 
\hline 
U-Net & 92.61  & 14.14  & 3.73  & 7.45 \tabularnewline
\hline 
U-Net with Augmentation & 94.64 & 7.89 & 2.66 & 3.32 \tabularnewline
\hline 
HC18 current best  & \textbf{98.10}  & \textbf{0.93 } & \textbf{1.17}  & \textbf{0.64 }\tabularnewline
\hline\hline 
SCU-Net  & 89.04  & 11.23  & 6.46 & 7.31 \tabularnewline
\hline 
Late Fusion  & 95.37  & 3.72  & 2.36  & 1.93 \tabularnewline
\hline 
\begin{tabular}{@{}c@{}}
DU-Net\tabularnewline
\end{tabular} &  97.33  & 1.41  & 1.58  & 0.97  \tabularnewline
\hline 
\end{tabular}} 

\caption{\label{tab:HCperformance} Performance on Fetal Head Segmentation}
\end{table}
The Dice coefficient (DC) and Hausdorff distance (HD), in mm, between the GT and obtained HC on the test set, are presented in Table \ref{tab:HCperformance}.  \textit{DUNet} shows a significant improvement over the baseline UNet. However, the performance is lower than the top results in HC18. This is partly due to the errors introduced in ellipse fitting to our results as the assessment is done in terms of ellipse parameters. It is also noteworthy that the top-performing methods have used complex networks with attention, dilated convolutions, etc. Since our goal is to demonstrate the effectiveness of hybrid representations as a proof of concept, the network was deliberately kept simple. Thus, the performance of \textit{DU-Net} is promising. Fig \ref{fig:US} shows the qualitative results on a few samples from the validation set (due to the unavailability of ground truth for test images). We found that the U-Net over-segmented in most of the images. 

\section{Conclusion}
We argued for a hybrid approach to segmentation with deep neural networks. Scattering Coefficients were used as additional input features for U-net based image segmentation. The proposed \textit{DU-Net} performed better than (for OD and  OC segmentation) or close to (for HC computation) state of the art despite no augmentation and use of a simpler network. The performance boost over the U-net was significant. The proposed method offers a way to dispense with the need for data augmentation, which increases the training time. It also motivates us to explore the use of other handcrafted features (wavelets or others) in place of SCs. 

\bibliographystyle{IEEEbib}
\bibliography{strings}
\end{document}